\documentclass[12pt]{article}
\usepackage[utf8]{inputenc}

\usepackage[normalem]{ulem}
\usepackage[permil]{overpic}
\usepackage{tikz}
\usepackage{mathtools}
\usepackage{amsmath}
\usepackage{graphicx}% Include figure files
\usepackage{dcolumn}% Align table columns on decimal point
\usepackage{bm}% bold math
\usepackage{color}
\usepackage[normalem]{ulem}
\usepackage{hyperref}

\def\txtd{{\textnormal{d}}}

\def\I{\infty}

\newcommand{\be}{\begin{equation}}
\newcommand{\ee}{\end{equation}}
\newcommand{\benn}{\begin{equation*}}
\newcommand{\eenn}{\end{equation*}}
\newcommand{\bea}{\begin{eqnarray}}
\newcommand{\eea}{\end{eqnarray}}
\newcommand{\beann}{\begin{eqnarray*}}
\newcommand{\eeann}{\end{eqnarray*}}

\usepackage{amsfonts,epsfig,amsmath,amsthm,amssymb,graphics,verbatim}
\usepackage[ margin=1in]{geometry}

\title{Balancing quarantine and self-distancing measures in adaptive epidemic networks}

\author{Leonhard Horstmeyer$^{1,2}$~, Christian Kuehn$^{1,2,3}$~, and Stefan Thurner$^{2,4,5}$}

\footnotetext[1]{equal contribution}
\footnotetext[2]{Complexity Science Hub Vienna, Josefst\"adter Str.~39, 1080 Vienna, Austria.}
\footnotetext[3]{Technical University of Munich, Department of Mathematics (M8), Boltzmannstr.~3, 85748 Garching b.~M\"unchen, Germany.}
\footnotetext[4]{Section for Science of Complex Systems, Medical University of Vienna, Spitalgasse 23, 1090 Vienna, Austria.}
\footnotetext[5]{Santa Fe Institute, 1399 Hyde Park Road, Santa Fe, NM 85701, USA.}

\date{}

\begin{document}

\maketitle

\begin{abstract}
% ST Ein bisschen gestrafft 
%In this work, 
We study the
%are interested in 
%
%the balance of % or 
relative importance of
two key control measures for epidemic spreading: endogenous social self-distancing and exogenous imposed quarantine. We use the framework of adaptive networks, moment-closure, and ordinary differential equations (ODEs) to 
%build and study 
introduce several novel models based upon susceptible-infected-recovered (SIR) dynamics. First, we compare 
%full, 
computationally 
%very 
expensive, adaptive network simulations with 
%smaller-scale, 
their corresponding 
computationally 
highly efficient 
%very fast, 
ODE  
equivalents 
and find excellent agreement. Second, we discover that there 
%is 
exists
a relatively simple critical curve in parameter space for the epidemic threshold, which strongly suggests 
%that there is 
a mutual 
compensation effect between the two mitigation strategies: %More precisely, 
as long as 
%the 
social distancing and quarantine measures are both 
sufficiently strong,  
%in tandem, then 
large outbreaks 
%can be 
are prevented. Third, we 
%proceed to also 
study the total number of infected and the maximum peak during large outbreaks using a combination of analytical estimates and numerical simulations. Also 
for 
%the study of 
large outbreaks 
%shows 
we find
a similar compensation effect as for the epidemic threshold. This suggests that if there is 
%very 
little incentive for social distancing 
within 
%internal to 
a population, 
%extremely 
drastic quarantining is required, and vice versa. Both 
pure 
scenarios are 
%practically 
unrealistic 
in practice. 
%so 
Our models show that only a combination of measures is likely to succeed to 
%stop and 
control epidemic spreading. Fourth, 
we 
analytically
compute
%develop novel 
an upper bound 
 %techniques 
for 
%the analytical studies of 
the total number of infected on 
adaptive 
networks,
using 
integral estimates in combination with the 
%application of 
moment-closure approximation on the level of an observable. This is a 
%major 
methodological innovation. 
%Furthermore, 
%Our general modelling shows,
Our method allows us to elegantly and quickly check and cross-validate 
%many 
various conjectures about the relevance of different network control measures.
%can be checked and cross-validated 
%quite 
%elegantly and quickly, which may make 
In this sense 
it becomes 
possible to adapt models  
%more 
rapidly to new epidemic 
%threads 
challenges 
such as the recent COVID-19 pandemic.  
\end{abstract}

\newpage
%%%%%%%%%%%%%%%%%%%%%%%%%%%%%%%%%%%%%%%%%%%%%%%%%%%%%%%%%%%%%%%%%%%%%%%%%%%%%%%%%
\section{Introduction}
\label{sec:intro}

The recent COVID-19 pandemic has demonstrated the necessity to develop and study effective models of epidemic dynamics~\cite{AndersonMay}. Classical epidemic models are compartmental models leading to relatively simple low-dimensional ordinary differential equations (ODEs)~\cite{BrauervandenDriesscheWu,DiekmannHeesterbeek}. These minimal ODE models can be derived from first principles~\cite{KissMillerSimon} but often suffer from very strong assumptions, such as a sufficiently high link density within the network of individuals. Within the past two decades, it became apparent that viewing the structure of the contagion process via a network science approach is crucial~\cite{Pastor-SatorrasVespignani,ColizzaBarratBarthelemyVespignani,Durrett,HouseKeeling,Pastor-Satorrasetal,ThurnerKlimekHanel}. During the COVID-19 pandemic it became clear that there are two major effects controlling direct epidemic spreading in humans without an available vaccine or immediate medical treatment: exogenous quarantine measures~\cite{MaierBrockmann,Kucharskietal} and endogenous social self-distancing (or social avoidance) of existing contacts~\cite{Giordanoetal}. Formally, measures can be considered on a finer scale, such as (digital) contact tracing~\cite{Ferrettietal,Kretzschmaretal}. Yet, most non-pharmaceutical interventions (NPI) can be grouped into external/exogenous ones leading effectively to quarantine-type effects, and intrinsic/endogenous ones within a population that lead effectively to a social re-organization of contact networks. 

In this work, we are interested in developing and analyzing effective, yet tractable mathematical network epidemic models to understand how to compare and balance the effects of quarantining and social self-distancing. Motivated by the COVID-19 pandemic~\cite{ThurnerKlimekHanel}, we start from a standard susceptible-infected-recovered (SIR) model on a complex network~\cite{KissMillerSimon}. Next, we use two well-established modelling approaches. First, we simply add another possible quarantine state, $X$, of the nodes~\cite{MaierBrockmann,Peaketal} together with a transition rate, $\kappa$, of infected individuals to transition to state $X$. Second, we use a social self-distancing rule of susceptible individuals trying to avoid contact with infected individuals leading to re-wiring of links~\cite{GrossDLimaBlasius,ShawSchwartz,Risau-GusmanZanette} controlled by a re-wiring rate, $w$; we keep the population size and the total number of links fixed to account for the propensity to keep social contact. Note that re-wiring links makes the network fully adaptive~\cite{GrossSayama}, i.e., there is dynamics on and of the network. The resulting model is a Markov chain on all possible node states and all possible edge configurations. It can be simulated on small to medium size networks, but becomes quickly computationally intractable on large networks. For this reason, we derive suitable ODE approximations~\cite{KissMillerSimon} based on a moment-closure approximation~\cite{Keeling,KeelingRandMorris,GrossDLimaBlasius,KuehnMC}. This approximation technique can also account for the dynamically changing connectivity of the network. We obtain a hierarchy of models. A particular model is obtained after fixing a truncation level. In our analysis we focus on the second-order or pair-approximation moment-closure, which leads to a five-dimensional ODE. We compare medium-size direct network simulations with ODE simulations. 

As a next step, we investigate the main questions associated with SIR-type models: (I)
Does an epidemic spread happen, or does it  decay immediately? (II) How big is the cumulative size of the epidemic outbreak? (III) What is the maximum size of the infected population during an epidemic? As expected, the first question (I) can be calculated directly using local analysis and we can express the epidemic threshold as a function of the re-wiring rate $w$ and the quarantine transition rate $\kappa$ in the approximating ODEs. Questions (II)-(III) are much harder to address as network structure effects preclude the application of classical methods from mathematical epidemiology to calculate the SIR outbreak size and/or the maximum peak. Here we develop a new technical tool by viewing the outbreak size as a global observable and applying moment-closure methods and integral estimates on the level of this observable. This technique leads to an upper bound on the global outbreak size within suitable parameter regimes, which we cross-validate numerically. 

From a public health perspective, we find that large parameter regimes in the $(w,\kappa)$-plane show a linear, or almost linear relationship regarding the effects of quarantine versus social self-distancing leading to a bounded triangular region, within which the epidemic cannot be avoided, controlled, or contained efficiently. This shows that there is a balancing effect between strong quarantining and social self-distancing, i.e., the weakening of one measure necessitates the strengthening the other and vice versa. On the one hand, this is an intuitive result. On the other hand, it arises without any major assumptions in broad parameter regimes in a complex fully-adaptive network epidemic model. In addition, it can be observed in network- and ODE simulations, and can be analytically treated both, locally and globally via nonlinear dynamics techniques. Hence, it seems advisable, when reducing or observing the reduction of mitigation measures, i.e., a decrease of $w$ and/or $\kappa$, to avoid entering the dangerous triangle region as there is no easy way to leave it again
by a simple small change.\medskip  

The paper is structured as follows: In Section~\ref{sec:model} we provide the detailed mathematical model for adaptive social self-distancing models with quarantine including moment-closure via pair approximation; in the appendix we also develop more complicated higher-order moment-closure models. In Section~\ref{sec:results} we present our main analytical and numerical results including a new global moment-closure viewpoint for global observables. In Section~\ref{sec:conclusion}, we provide a summary and outlook, how our approach can be extended to wider set of applications. 

%%%%%%%%%%%%%%%%%%%%%%%%%%%%%%%%%%%%%%%%%%%%%%%%%%%%%%%%%%%%%%%%%%%%%%%%%%%%%%%%%
\section{Adaptive SIR-type Network Models}
\label{sec:model}

Here we compare two different, yet comparable, measures present in most epidemics: social self-distancing, i.e., nodes/agents avoid infected individuals simply due to the risk of acquiring the disease themselves, and external quarantine measures, which enforce the removal of infected nodes from the population. It is evident from data that both measures have played a key role during the COVID-19 pandemic~\cite{MaierBrockmann,Kucharskietal,Giordanoetal}. To account for the complex social structure, we start with microscopic Markov process  models of susceptible-infected-recovered (SIR) dynamics on general networks with $N$ nodes, $K$ undirected links, and node states $S$, $I$, and $R$. Then we add mitigation measures to the SIR model. The well-known basic SIR rules are:

\begin{itemize}
 \item (infection) infected $I$ nodes infect susceptible $S$ nodes along susceptible-infected $SI$ links with a rate $\beta>0$.
 \item (recovery/death) infected $I$ nodes become recovered $R$ nodes at a rate $\gamma>0$. 
\end{itemize}
 
One way to model social self-distancing~\cite{GrossDLimaBlasius,ShawSchwartz} is the preference of the susceptible $S$ nodes to avoid interactions with the infected $I$ nodes:

\begin{itemize}
 \item (social self-distancing) $SI$ links are re-wired to susceptible-susceptible $SS$ links at a rate $w\geq 0$.
\end{itemize}

The self-distancing/re-wiring rule makes the network fully adaptive~\cite{GrossSayama} and allows for very general network topologies. The rule also takes into account that links are not lost, which mirrors the desire to keep as many social connections as possible and to optimally re-wire them to mitigate risk. While it is straightforward to simulate the resulting Markov process on any given network, the simulations become prohibitively expensive for large $N$. It is also straightforward to use the master equation for the resulting Markov process~\cite{Norris} and arrive at the following set of ODEs via standard techniques~\cite{KissMillerSimon}
\begin{equation}
\label{eq:SIRadaptive}
    \begin{array}{lclcl}
 \dot{[S]}&=&\frac{\txtd }{\txtd t} [S]
 &=&
 -\beta [SI],
 \\
 \dot{[I]}&=&\frac{\txtd }{\txtd t} [I]
 &=&
 \beta [SI] - \gamma [I],
 \\
 \dot{[R]}&=&\frac{\txtd }{\txtd t} [R]
 &=&
\gamma [I],
 \\
\dot{[SI]}&=& \frac{\txtd }{\txtd t} [SI]
 &=&
 -(\beta+\gamma+w)[SI] + \beta[SSI] -\beta[ISI], 
 \\
 \dot{[SS]}&=&\frac{\txtd }{\txtd t} [SS]
 &=&
- \beta [SSI] + w\frac{[S]}{[R]+[S]} [SI],
\end{array}
\end{equation}
where $[S]=[S](t)$, $[I]=[I](t)$, $[R]=[R](t)$, $[SI]=[SI](t)$, $[SS]=[SS](t)$, $[SSI]=[SSI](t)$ and $[ISI]=[ISI](t)$ are expectation values of the number of susceptible, infected and recovered nodes, of $SI$-links and $SS$-links and of $SSI$- and $ISI$-triplet motifs. The ordinary differential equations (ODEs)~\eqref{eq:SIRadaptive} represent a variation of earlier models~\cite{ShawSchwartz,GrossDLimaBlasius}. Note that we do not allow recovered individuals to pass back into the susceptible compartment, which makes sense if we assume that immunity is acquired upon recovery, e.g., as conjectured to be true on certain time scales of interest for COVID-19. Although the ODEs~\eqref{eq:SIRadaptive} are actually exact in the mean-field limit~\cite{KissMillerSimon} for any graph, they are not closed as we have not written down the equations for the $SSI$ and $ISI$ motifs. Although these equations could be derived, they would depend on fourth-order motifs, and so on~\cite{KuehnMC,HouseKeeling}. To avoid studying an infinite system of ODEs, we employ a standard moment-closure pair approximation~\cite{KeelingRandMorris,Keeling,KissMillerSimon,GrossDLimaBlasius}, assuming that 
\benn
[ABC]\approx m(A,B)m(B,C)\frac{[AB][BC]}{[B]}\qquad \text{for }A,B,C\in\{S,I,R\},
\eenn
where $m(A,B) =2$ if $A=B$ and $m(A,B) =1$ if $A\neq B$. With this closure, one obtains a system of four ODEs for the densities 
%$\rho_I=\frac{[I]}{N}$ 
$\rho_I=[I]/ N$ 
and 
%$\rho_R=\frac{[R]}{N}$ 
$\rho_R=[R]/ N$ 
of infected and recovered nodes and the per-node densities of susceptible-infected and susceptible-susceptible links:
 \begin{equation}
 \label{eq:SIRsimple}
 \begin{array}{lcl}
 \frac{\txtd }{\txtd t} \rho_I 
  &=&
   \beta \rho_{SI} - \gamma \rho_I,
   \\
   \frac{\txtd }{\txtd t} \rho_R 
  &=&
    \gamma \rho_I,
\\
   \frac{\txtd }{\txtd t} \rho_{SI}
&=&
-(\beta+\gamma+w) \rho_{SI} +\beta \rho_{SI}\frac{ 2 \rho_{SS}-\rho_{SI}  }{1-\rho_I-\rho_R},
\\
  \frac{\txtd }{\txtd t} \rho_{SS}
&=&
-2\beta \frac{\rho_{SI} \rho_{SS}}{1-\rho_I-\rho_R}
+w\left[\frac{1-\rho_I-\rho_R}{1-\rho_{I}}\right]\rho_{SI}.
\end{array}
 \end{equation}
Here we made use of node conservation
and the notation 
%$\rho_{AB}\approx \frac{[AB]}{N}$ 
$\rho_{AB}\approx [AB]/N$ 
to emphasize that we are working with approximate per-node densities after moment-closure has been applied. As such, the equations~\eqref{eq:SIRsimple} only cover the aspect of social self-distancing and take into account the complex adaptive network structure via a second-order closure.

\begin{figure}
    \centering
    \begin{overpic}[width=0.48\linewidth]{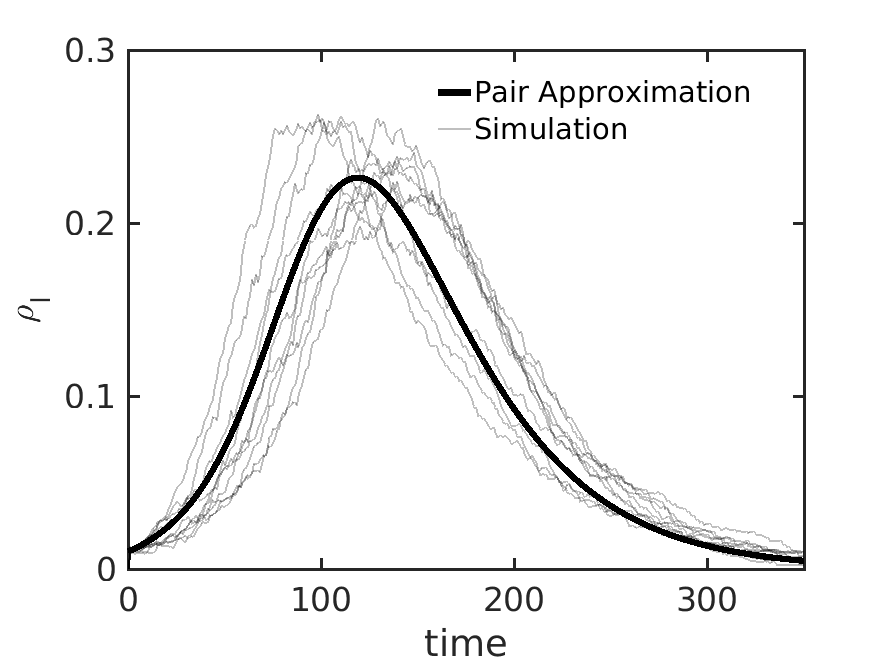}%
    \put(0,700){a)}%
    \end{overpic}
    \begin{overpic}[width=0.48\linewidth]{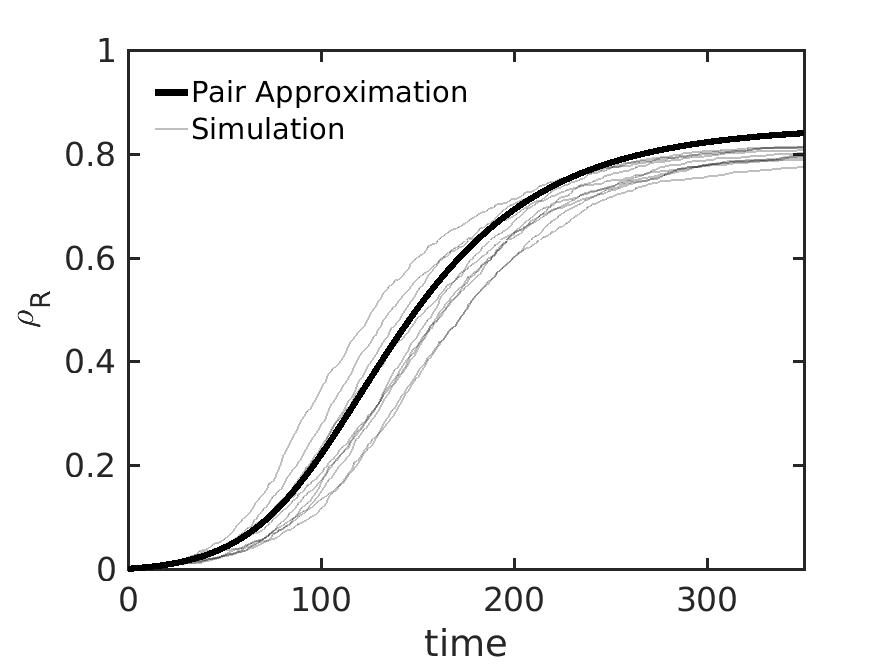}%
    \put(0,700){b)}%
    \end{overpic}
    \caption{Sample paths for the adaptive SIRX model (thin line) and the Pair Approximation from \eqref{eq:mfa} (thick line). 
    In (a) we depict the disease prevalence ($\rho_I$) and in (b) we depict the cumulative size of the recovered compartment ($\rho_R$). The dynamical parameters are given by a recovery rate of $\gamma=0.025$ and an infection rate of $\beta=0.005$. The intervention parameters for the quarantine and re-wiring rates are $\kappa = w = 0.0025$. The release rate from the quarantined compartment is $\delta = 0.001$. For the simulation we sampled from an Erd\~os-Rényi ensemble of size $N=2000$ with mean degree $\mu=15$. We initialized $1\%$ of nodes as infected $\rho_I(0) = 0.01$ and $\rho_{SI}(0) = \mu \rho_I(0)$. 
    }
    \label{fig:1}
\end{figure}

We take into account quarantine effects, such as in the modelling of COVID-19 in~\cite{MaierBrockmann}. In~\cite{MaierBrockmann}, network structure was not considered. Quarantine effects lead to certain features of epidemic spreading that cannot be captured by classical SIR models. We denote the quarantined compartment by $X$. The rules we use are:

\begin{itemize}
 \item (quarantine) infected $I$ nodes are quarantined into a state $X$ at a rate $\kappa\geq 0$.
 \item (recovery of $X$) quarantined nodes are released into the recovered compartment $R$ at rate $\delta>0$.
\end{itemize}

In particular, we consider quarantining and rewiring only for the infected compartment. This is a simplification of the present model and is not pursued like this in many real contact tracing efforts. The expected release time from the $X$-compartment is $\langle T\rangle = 1/\delta$, since the rates are Poissonian. However, $\delta$ does not have an effect on the amount of nodes in the infected compartment in this model and for any positive $\delta$ the amount of nodes in the recovered compartment for $t\to\infty$ is also independent of $\delta$. With these quarantine rules, we obtain the non-closed moment equations:
\begin{equation}
\label{eq:adaptiveSIRX}
\begin{array}{lcl}
 \frac{\txtd }{\txtd t} [S]
 &=&
 -\beta [SI], 
 \\
 \frac{\txtd }{\txtd t} [I]
 &=&
 \beta [SI] - (\gamma + \kappa)[I], 
 \\
 \frac{\txtd }{\txtd t} [R]
 &=&
\gamma [I] +\delta[X],
 \\
 \frac{\txtd }{\txtd t} [X]
 &=&
\kappa[I] - \delta[X],
 \\
 \frac{\txtd }{\txtd t} [SI]
 &=&
 -(\beta+\gamma+w)[SI] + \beta[SSI] -\beta[ISI] - \kappa [SI],
 \\
 \frac{\txtd }{\txtd t} [SS]
 &=&
- \beta [SSI] + w\frac{[S]}{[R]+[S]} [SI].
\end{array}
 \end{equation}
 More complicated variants of the rules are discussed in Appendix~\ref{appendixA}. 
 We see that the parameters appear linearly in the equations, so that any one of them can be used to re-scale the time, e.g. $t\mapsto \gamma t$. This leaves four effective dynamical parameters $\beta, w, \kappa$ and $\delta$. The last one does not affect the infected compartment or the recovered compartment at infinity and is therefore not part of the subsequent analysis.
 
 Using a moment-closure pair approximation, we get the closed system 
\begin{equation}
\label{eq:mfa}
  \begin{array}{lcl}
  \frac{\txtd }{\txtd t}{\rho_S}&=&-\beta \rho_{SI},\\ 
  \frac{\txtd }{\txtd t}{\rho_{I}}&= &
  \beta\rho_{SI}-(\kappa+\gamma)\rho_{I},\\
  \frac{\txtd }{\txtd t}{\rho_{R}}&=&
  \gamma\rho_I +\delta (1-\rho_{S}-\rho_{I}-\rho_{R}),\\
  \frac{\txtd }{\txtd t}{\rho_{SI}}&=&
  -(\beta+\gamma+w+\kappa)\rho_{SI}+\beta\rho_{SI}\frac{2\rho_{SS}-\rho_{SI}}{\rho_S},\\
  \frac{\txtd }{\txtd t}{\rho_{SS}}&=& - 2\beta\frac{\rho_{SI}\rho_{SS}}{\rho_{S}}+w\frac{\rho_S}{\rho_S+\rho_{R}}\rho_{SI},
  \end{array}
 \end{equation}
 that allows us to compare the effects of social self-distancing and quarantine. We compare this model with full network simulations in Figure~\ref{fig:1}. The results show excellent agreement for the vast majority of sample runs for a large part of the parameter space, when $w,\kappa>0$; see also Appendix~\ref{sec:apB} for additional comparisons, where even for the singular cases $w=0$ or $\kappa=0$ excellent agreement is observed. 

%%%%%%%%%%%%%%%%%%%%%%%%%%%%%%%%%%%%%%%%%%%%%%%%%%%%%%%%%%%%%%%%%%%%%%%%%%%%%%%%%
\section{Results}
\label{sec:results}

In contrast to SIS or SIRS, an epidemic eventually dies out for a standard SIR model. Hence, three questions arise:
\begin{enumerate}
    \item[(I)] Given an initial density of infected $I(0)$ sufficiently close to the disease-free state, does the epidemic spread, or does it die out almost immediately?
\item[(II)]How big is the cumulative size of the epidemic outbreak $R_\infty$ (we use $r_\infty$ for the corresponding density)? $R_\infty$ measures the total number of nodes at $t\to\infty$ in the recovered compartment $R$ (respectively $r_\infty$ is the corresponding density).
\item[(III)]What is the maximum size of the epidemic, $\hat{[I]}:=\max_tI(t)$, i.e., what is the height of the highest peak?
\end{enumerate}\medskip  

To answer (I), the local calculation near the disease-free state is relatively simple if we have a closed ODE model. For example, consider the adaptive SIR model without quarantining~\eqref{eq:SIRsimple} and use the disease-free state $\rho_*$ with
\begin{equation*}
\rho_I=\rho_R=\rho_{SI}=0\quad \text{and} \quad \rho_{SS}=\mu/2,
\end{equation*}
i.e., also all links are of type $SS$. Here $\mu$ is the average degree of the network, which equals 
%$\frac{2K}{N}$ 
$2K/N$ 
for a network with $K$ edges. Linearizing the vector field at $\rho_*$, we find that for 
\begin{equation}
    \beta<\beta^{\textnormal{adp}}_c=\frac{\gamma+w}{\mu-1}, 
\end{equation}
an epidemic dies out exponentially fast. A very similar calculation for the full model \eqref{eq:mfa} reveals:
\begin{equation}
    \beta<\beta^{\textnormal{qurt+adp}}_c=\frac{\gamma+w+\kappa}{\mu-1}, \label{eq:critcurve}
\end{equation}
as the critical threshold for the infection rate. We see that on a local level near $\rho_*$, the effects of self-distancing and quarantine are comparable as the rates of both processes lower the critical threshold in a linear way.

\begin{figure}
    \centering
    \begin{overpic}[width=0.495\linewidth]{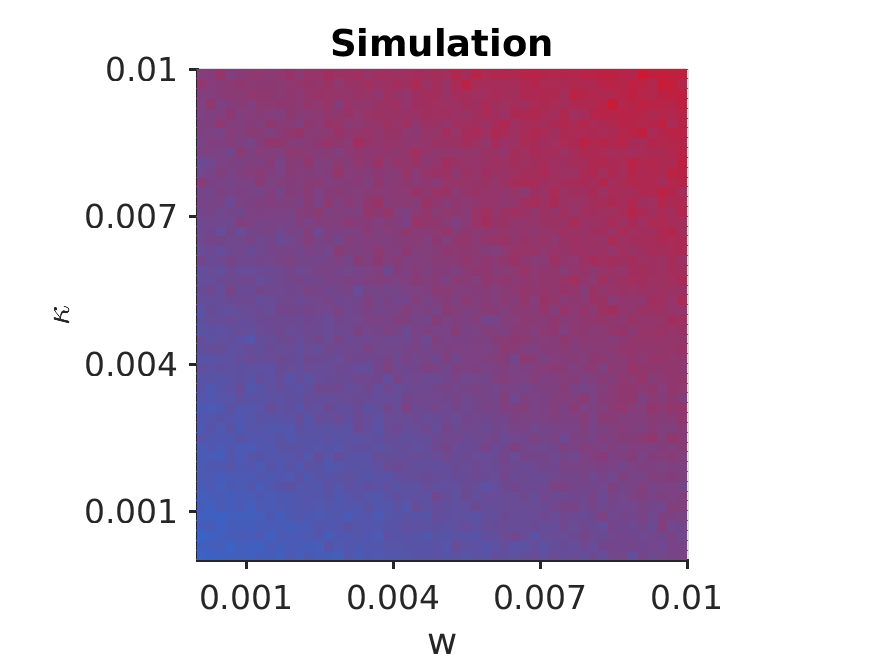}%
    \put(0,700){a)}%
    \end{overpic}
    \begin{overpic}[width=0.495\linewidth]{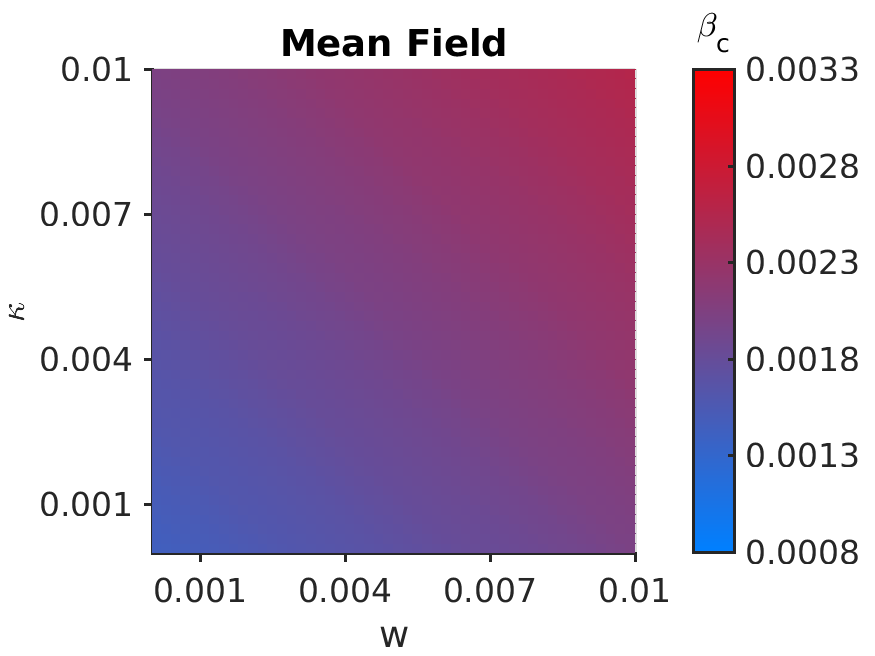}%
    \put(0,700){b)}%
    \end{overpic}
    \caption{Comparison of the critical infection rate. We depict the rate $\beta$ at which the epidemics surpasses a threshold of $r_\infty=0.05$, which we take as a proxy for the critical point $\beta_c$. We compare the simulations (left) with the mean-field analysis from the pair approximation (right). The parameters are as before, except for the size $N=500$.}
    \label{fig:2}
\end{figure}

 These results are completely consistent with the numerics. Figure~\ref{fig:2} shows the rate $\beta$ at which we record an epidemic outbreak/growth for direct network simulation, as well as for numerical integration of the mean-field ODEs. The linear level-set structured in the $(w,\kappa)$-diagram expected from \eqref{eq:critcurve} is clearly visible on the network simulation and the ODE integration levels. This answers questions (I), and means that a combination of measures is particularly effective to contain a disease early on. Since it is unrealistic to assume that social self-distancing happens effectively in the situation of a new disease, our SIRX model suggests that one has to compensate and focus more on quarantine measures of infected individuals in the early phase.\medskip

\begin{figure}
    \centering
    \begin{overpic}[width=0.495\linewidth]{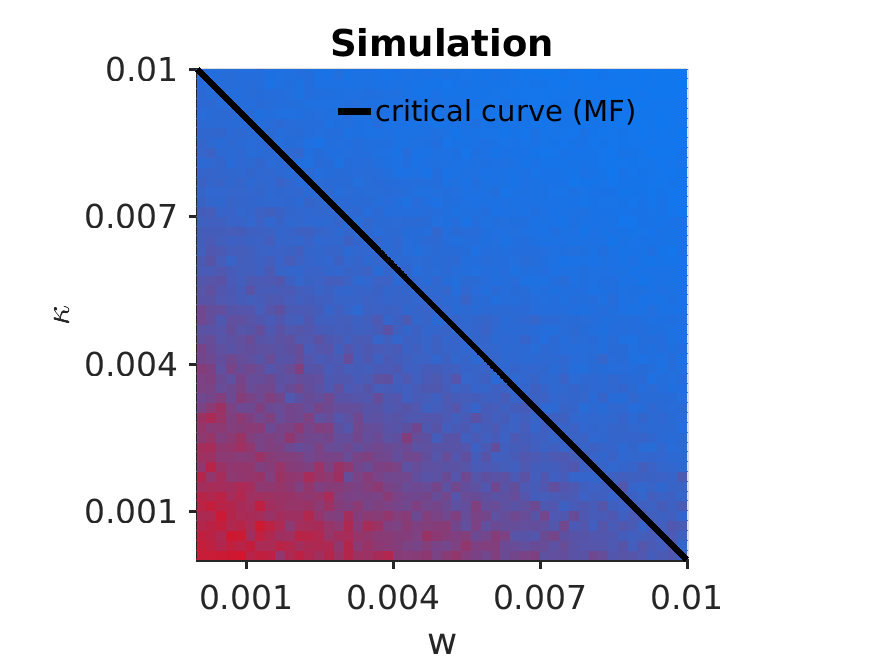}%
    \put(0,700){a)}%
    \end{overpic}
    \begin{overpic}[width=0.495\linewidth]{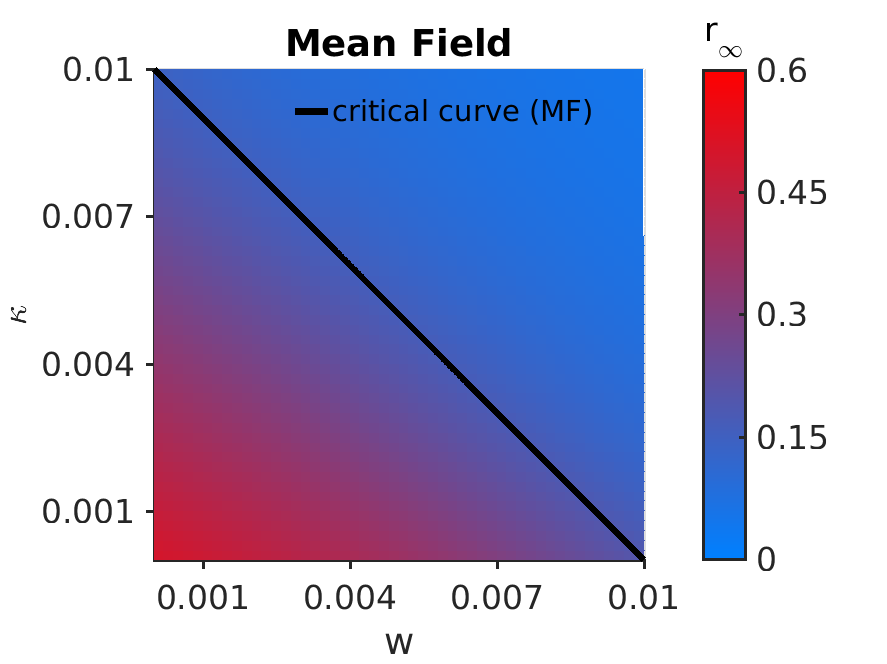}%
    \put(0,700){b)}%
    \end{overpic}
    \caption{Comparison of the overall size of the epidemic  $r_\infty$ for a fixed $\beta=0.0025$ in the $(\kappa,w)$-parameter plane. (a) Simulations. (b) Mean-field analysis from the pair approximation. We also indicate the critical curve, as calculated from \eqref{eq:critcurve}. Again $N=500, \mu=15, \gamma=0.025$, $\delta=0.01$ and $\rho_I(0)=0.01$.}
    \label{fig:3}
\end{figure}

However, the local structure near the disease-free state only yields a partial picture of SIR models. In fact, one often does observe epidemic outbreaks for SIR-type dynamics. For this case, we study $r_\infty$ to answer our second question (II). Figure~\ref{fig:3} shows $r_\infty$ for a range of values near the epidemic transition in the $(w,\kappa)$-plane. We compare numerical simulations that estimate $\langle \lim_{t\to\infty}R(t)/N\rangle$ with the pair approximation $\rho_{R,\I}:=\lim_{t\to\infty}\rho_R(t)$ of equation \eqref{eq:mfa}.

The numerical results indicate another linear relation between the parameters $w$ and $\kappa$. In general, it is impossible to get an exact formula for the cumulative size of the epidemic outbreak for an arbitrary model for SIR-type dynamics on complex networks. Yet, we can arrive at an implicit formula starting with in our adaptive SIRX model \eqref{eq:adaptiveSIRX}. We denote the expected final number of recovered individuals by $R_\I:=\lim_{t\to\infty}[R](t)$ and write it as
\begin{equation*}
    R_\I=R_\I-[R](0)=\int_0^\infty \dot{[R]}~\txtd t,
\end{equation*}
where we used $[R](0)=0$ and we have omitted the argument $t$ of the last integrand for brevity. Now we use the differential equation for $[R]$ and insert it to get 
\begin{equation*}
    R_\I=\gamma \int_0^\infty  [I]~\txtd t + \delta\int_0^\I [X]~\txtd t.
\end{equation*}
We have obtained two integrals, which would suffice to calculate $r_\I$. Using the same idea as for $[R]$, we find for $[I]$ and $[X]$ that
\begin{eqnarray*}
    0-[I](0)&=&\int_0^\infty  \dot{[I]}~\txtd t \;=\,\beta\int_0^\infty[SI]~\txtd t - (\kappa+\gamma)\int_0^\infty[I]~\txtd t\\
    0&=&\int_0^\infty  \dot{[X]}~\txtd t ~ \;=\, \kappa\int_0^\infty[I]~\txtd t -\delta\int_0^\infty[X]~\txtd t,
\end{eqnarray*}
as there cannot be any quarantined infected nodes in the beginning or at the end of the epidemic. Using the ODE for $[I]$, we get
\begin{equation*}
    R_\I=\gamma\int_0^\infty [I] ~\txtd t +\kappa\int_0^\infty [I] ~\txtd t = \beta \int_0^\infty [SI]~\txtd t + [I](0).
\end{equation*}
Several crucial observations are evident from the last formula. The procedure generically does not terminate on the infinite network level as in generic cases, we expect that all motifs could eventually occur. This means that without further assumptions only an infinite series expression for $R_\I$ is obtained or we could obtain upper or lower bounds. Still, the infinite series and particularly upper bounds are very informative as they can display the influence of the different parameters  and link them to higher-order network motifs. Indeed, at the next step, the expected number of links comes into play. We get
\begin{equation*}
0-[SI](0)=\int_0^\I \dot{[SI]}~\txtd t=
 -(\beta+\gamma+w+\kappa)\int_0^\I [SI]~\txtd t + \beta\int_0^\I[SSI]~\txtd t -\beta\int_0^\I [ISI]~\txtd t,
 \end{equation*}
and we obtain 
\begin{equation}
R_\I = [I](0) + \frac{\beta}{\beta+\gamma+w+\kappa}[SI](0) + \frac{\beta^2}{\beta+\gamma+w+\kappa}\int_0^\infty\Big(
    [SSI] - [ISI]\Big) ~\txtd t.
\label{eq:infiniteseries}
\end{equation}
The infinite sum will yield new motif terms involving infected nodes at every step at the time $t=0$. This demonstrates the key importance of the network structure. For example, a highly connected first cluster of infected nodes yields a large number of $SI$-links and thereby a large final outbreak size. We could continue this procedure to obtain an infinite series formally but this does not give any concrete quantitative approximations. Instead, we aim for an upper bound of the total number of infected/recovered. 

\begin{figure}
    \centering
    \begin{overpic}[width=0.48\linewidth]{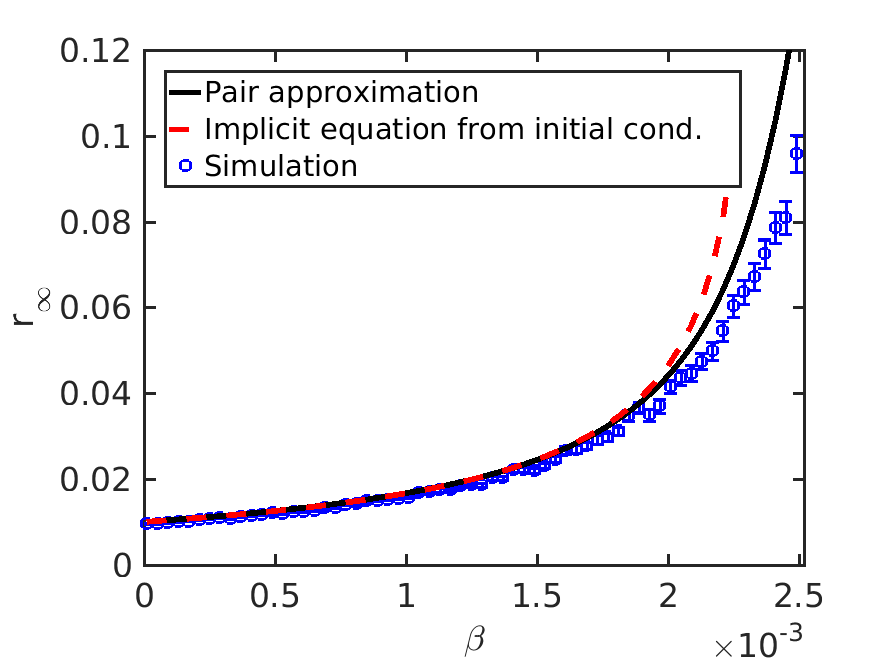}%
    \put(0,700){a)}%
    \end{overpic}
    \begin{overpic}[width=0.48\linewidth]{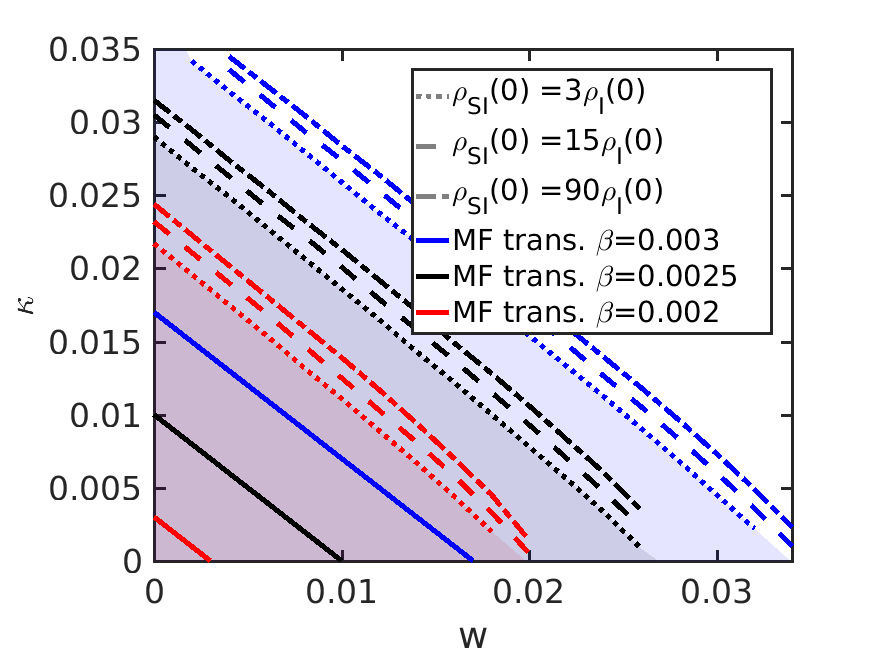}%
    \put(0,700){b)}%
    \end{overpic}
    \caption{Approximations of $r_\infty$. In (a) we show three approximations of $r_\I$ for the adaptive SIRX model via the Pair Approximation \eqref{eq:mfa}, the implicit equation \eqref{eq:boundrinf} and via repeated simulation of the stochastic dynamics. The implicit equation is an inequality in $r_\infty$ and depends on the initial conditions, which are here chosen in agreement with the other approximations, namely $\rho_I(0) = 0.01$ and $\rho_{SI}(0) = \mu \rho_I(0)$. The inequality is achieved by a H\"older bound, which requires a positivity condition on $\dot\rho_{SI}-2\dot\rho_{SS}\geq 0$ at all times. In  (b) we show the shaded regions where the positivity condition holds for a range of infection rate $\beta=0.003$ (blue, upper set of lines), $\beta=0.0025$ (black, middle set of lines) and $\beta=0.002$ (red, lower set of lines). For each infection rate we show the boundary for a set of initial conditions to illustrate the dependence on $\rho_I(0)$ and $\rho_{SI}(0)$. For the initial $SI$-link density we choose a mean-field scenario (dashed line) with $\rho_{SI}(0)=\mu \rho_{S}(0)$, a scenario (dash-dotted line) with dis-proportionally many initial $SI$-links $\rho_{SI}(0)=6\mu\rho_{S}(0)$ and a scenario (dotted) with very few $SI$-links $\rho_{SI}(0)=(\mu/5) \rho_{S}(0)$. One may also see the mean-field transition lines (solid lines) at the respective infection rates in the respective colours. They all lie within the positivity region. All parameters are as before, in particular $\mu=15$. }
    \label{fig:positivity}
\end{figure}

The key new technical step is that we directly impose the moment-closure pair approximation \emph{directly on} \eqref{eq:infiniteseries}, using the approximate densities $\rho_I, \rho_R, \rho_{SI}$ and $\rho_{SS}$ of the closed equations \eqref{eq:mfa}. We get an approximation for $r_\I$ in terms of $\rho_{R,\I}:=\lim_{t\to\I}\rho_R(t)$,
\begin{equation}
\label{eq:fbint}
    r_\I \approx \rho_{R,\I} = \rho_I(0) + \frac{\beta}{\beta+\gamma+w+\kappa}\rho_{SI}(0) + \frac{\beta}{\beta+\gamma+w+\kappa}\int_0^\infty\beta \frac{\rho_{SI}}{\rho_S}\Big(
    2\rho_{SS} - \rho_{SI}\Big) ~\txtd t.
\end{equation}
Using $\beta \rho_{SI}=-\dot{\rho}_{S}$ and applying the logarithmic derivative gives 
\begin{equation*}
    \rho_{R,\I}\approx  K_0 + \frac{\beta}{\beta+\gamma+w+\kappa}\int_0^\infty -\frac{\txtd}{\txtd t}(\ln(\rho_S))\Big(
    2\rho_{SS} - \rho_{SI}\Big) ~\txtd t,
\end{equation*}
where $K_0:=\rho_I(0) + \frac{\beta}{\beta+\gamma+w+\kappa}\rho_{SI}(0)>0$. Integration by parts and $\ln(x) = -|\ln(x)|$ for $x\in(0,1]$ yields
\beann
    \rho_{R,\I} &=& K_0 +
\frac{\beta}{\beta+\gamma+w+\kappa}\left(
- \left.\ln(\rho_S)\Big(
    2\rho_{SS} - \rho_{SI}\Big)\right|_{0}^{\infty}
    - 
\int_0^\infty |\ln(\rho_S)|\Big(
    2\dot{\rho}_{SS} - \dot{\rho}_{SI}\Big) ~\txtd t.
    \right)
\eeann    
Next, we use $|\ln(\rho_S(t))|\geq |\ln(\rho_{S,\I})| = -\ln(1-\rho_{R,\I})$ and we make the assumption that $2\dot\rho_{SS} - \dot\rho_{SI}\leq 0$ for all times (see the computations below for further justification of this assumption). Then we obtain the bound
\beann    
    \rho_{R,\I} &\leq &
    \;
K_0 +
\frac{\beta}{\beta+\gamma+w+\kappa}\left(
- \left.\ln(\rho_S)\Big(
    2\rho_{SS} - \rho_{SI}\Big)\right|_{0}^{\infty}
    + 
\ln\Big(1- \rho_{R,\I}\Big) \int_0^\infty \Big(
    2\dot{\rho}_{SS} - \dot{\rho}_{SI}\Big) ~\txtd t
    \right)\nonumber\\
&=    &
\;
K_0 +
\frac{\beta}{\beta+\gamma+w+\kappa}\bigg(
- \ln\Big(1-\rho_{R,\I}\Big)
    2\rho_{SS}(\infty) + \ln\Big(1-\rho_I(0)\Big)\Big(
    2\rho_{SS}(0) - \rho_{SI}(0)\Big)\\
&& \qquad\qquad\qquad\qquad\qquad\,
    + 
\ln\Big(1- \rho_{R,\I}\Big) \Big( 2\rho_{SS}(\I)-2\rho_{SS}(0) +\rho_{SI}(0)\Big)\bigg)\\
&=    &
\;
K_0 +
\frac{\beta}{\beta+\gamma+w+\kappa}\ln\left[
\frac{1-\rho_I(0)}{1-\rho_{R,\I}}
\right]\left(2\rho_{SS}(0) -\rho_{SI}(0)\right)
\eeann
Simplifying finally yields the desired upper bound
\be
\label{eq:boundrinf}
\rho_{R,\I}
\leq
\;
\rho_I(0) +
\frac{\beta}{\beta+\gamma+w+\kappa}\left[\rho_{SI}(0) + \left(2\rho_{SS}(0) -\rho_{SI}(0)\right)\ln\left[
\frac{1-\rho_I(0)}{1-\rho_{R,\I}}
\right]\right].
\ee 
The bound~\eqref{eq:boundrinf} is a transcendental inequality in $\rho_{R,\I}$. Regarding our assumption
\be
\label{eq:assumegrowth}
\dot\rho_{SI}\geq 2\dot\rho_{SS},
\ee
we find that it holds numerically for a broad ranges of parameters. In Figure~\ref{fig:positivity}(b) we show the domains of validity for the positivity assumption in the $(w,\kappa)$-plane for a range of infection rates and initial conditions. The assumption holds in a neighbourhood around the critical transition. 

Note that our analysis is in sharp contrast to the classical three-dimensional SIR ODE model, where an exact implicit functional relation for $r_\I$ can be obtained. Therefore, having an upper bound available such as~\eqref{eq:boundrinf} helps us to study the parameter dependencies. The same linear combination of the two parameters $w$ and $\kappa$ appears as in the local bifurcation case near the epidemic threshold. Now however, they occur via an inverse. The same conclusions as for the local epidemic spreading near the outbreak threshold are valid: we need a linear mix of quarantine and social self-distancing to keep the total number of infected $r_\I$ under control. Figure \ref{fig:positivity}(a) shows a comparison of the upper bound for  with numerical simulations of the full network as well as simulations of the pair approximation ODEs. Both capture the main trend well that occurs when the infection rate is increased. When the total infected population is around the ten percent level, our approximations show that employing a combination of quarantine and social-distancing might be effective in practice, while going beyond this level, a very steep increase of the total number of infected occurs.\medskip

\begin{figure}
    \centering
    \begin{overpic}[width=0.48\linewidth]{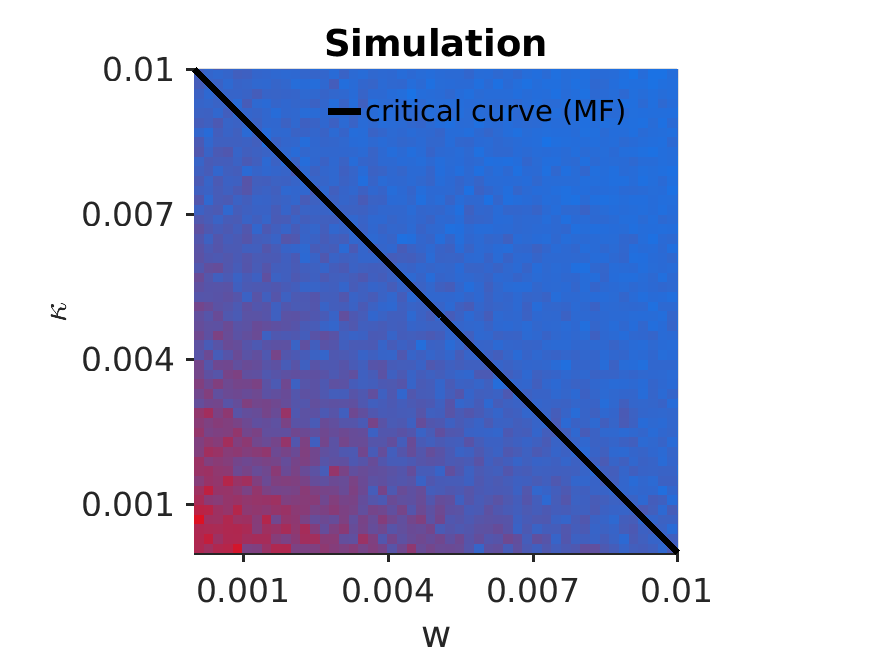}%
    \put(0,700){a)}%
    \end{overpic}
    \begin{overpic}[width=0.48\linewidth]{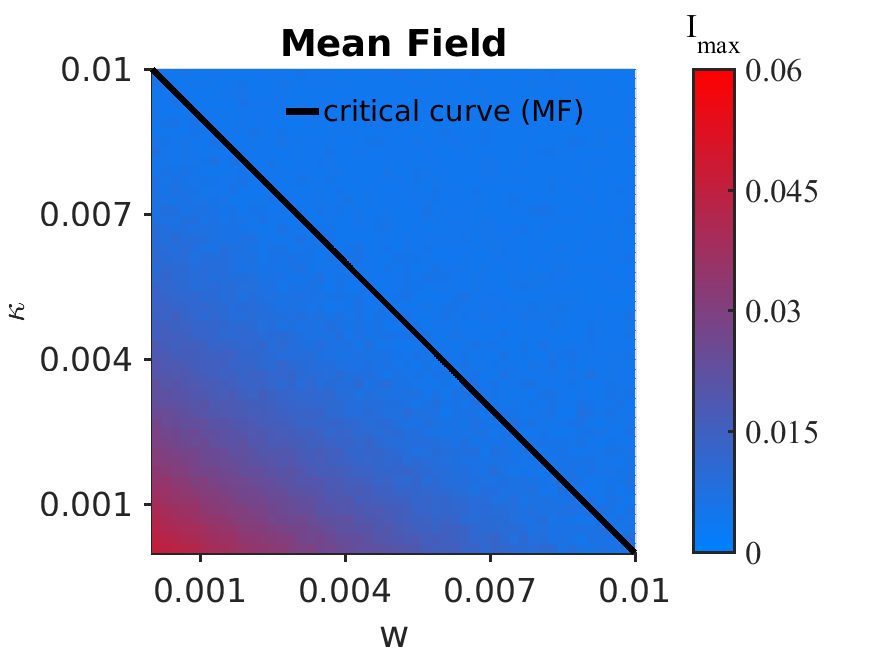}%
    \put(0,700){b)}%
    \end{overpic}
    \caption{Comparison of the maximal disease prevalence. (a) Simulation. (b) Mean-field from the Pair Approximation. We also indicate the critical curve, as calculated from \eqref{eq:critcurve}. We can see here a slight deviation, which can be explained by the fact that the simulations are random processes. The average of sample path maxima over many sample paths is not the same as the maximum of the average of sample paths. The former overestimates the expectation value of $\rho_I$. Again $N=500, \mu=15, \gamma=0.025, \beta=0.0025, \delta=0.01$, $\rho_I(0) = 0.01$ and $\rho_{SI}(0) = \mu \rho_I(0)$.}
    \label{fig:4}
\end{figure}

Finally, we answer question (III) regarding the maximum peak $\hat{[I]}$ of the number of infected, i.e., $\hat{[I]}/N$ is the maximal fraction across the entire duration of the epidemic. Figure~\ref{fig:4} shows the corresponding results comparing direct network simulations in Figure~\ref{fig:4}(a) with mean-field approximations in Figure~\ref{fig:4}(b). The structure of the results is familiar in the sense that a linear dependence between $w$ and $\kappa$ emerges for our studied parameter ranges. Therefore, one can conclude that using a \emph{well-tuned} combination of quarantine and social self-distancing outside of a triangular region in $(w,\kappa)$-space is likely to be not only effective in preventing outbreaks, or reducing the total number of infected during epidemic, but also to prevent high peaks.

\section{Conclusion \& Outlook}
\label{sec:conclusion}

In this work, we have provided three contributions. First, we developed a new type of adaptive network models that include two of the most important epidemic control measures: quarantine and social self-distancing. We derived mean-field models via pair approximation; even more detailed approximation schemes are discussed in the appendix. Second, we analyzed the new model via a numerical combination of direct network simulations and mean-field ODEs, which show excellent agreement. We focused on three questions regarding (I) the epidemic threshold, (II) the total number of infected individuals, and (III) the maximum peak of the epidemic. In all three cases, we demonstrated for a broad range that the parameters controlling quarantine and social-distancing enter in a comparable linear combination to control the epidemic spread. This has the practical implication that a suitable combination of these two measures outside of a well-defined triangular region in parameter space is the best choice as one cannot expect either measure to be executed perfectly in practice. Third, on a technical level, we have shown a new way to provide estimates for the total infected population during an epidemic by using pair approximation and integral estimates directly on the level of the final infected number observable. This provides a new technical tool for broad classes of epidemic models on networks since one can now aim to employ moment-closure on many other observables directly.\medskip

Of course, many generalizations of the presented model are possible. For example, one could try to use slightly different rules for the link dynamics allowing for link deletion~\cite{Balletal,TuncShkarayev}. Yet, we conjecture that the same analysis principles we have developed here still apply. Furthermore, it would be desirable to not only consider the mitigation of the epidemic itself but also whether quarantine and social self-distancing can help us or are detrimental to detect early-warning signs for large epidemic outbreaks~\cite{OReganDrake,WidderKuehn,Brettetal}. This line of research has already been started in recent years for epidemics on adaptive networks~\cite{KuehnZschalerGross,HorstmeyerKuehnThurner} but the interplay between pre-epidemic mitigation measures and warning signs has remained unexplored.\medskip 

\textbf{Acknowledgements:} CK acknowledges partial support by a Lichtenberg Professorship funded by the VolkswagenStiftung including the recently granted project add-on within the call ``Corona Crises and Beyond''. ST is grateful to the Austrian Research Promotion Agency (FFG) under projects 857136 and 873927, and the Medizinisch-Wissenschaftlicher Fonds des B\"urgermeisters der Bundeshauptstadt Wien under project CoVid004.

\bibliographystyle{alpha}
\bibliography{bib1}

%%%%%%%%%%%%%%%%%%%%%%%%%%%%%%%%%%%%%%%%%%%%%%%%%%%%%%%%%%%%%%%%%%%%%%%%%%%%%%%%%
\newpage
\appendix

\section{More on Moment Equations for Adaptive Epidemics with Quarantine}
\label{appendixA}

In this appendix, we provide more details on moment-closure schemes and the derivation of our reduced system. Beyond the pair approximation~\eqref{eq:mfa}, refined approximation schemes are possible, which we present in this appendix in more detail.

\subsection{Full closed equations up to second order motifs}

First, we write down the full closed equations up to second-order motifs, which are given by
\begin{align}
\frac{\txtd }{\txtd t} [S]
 =&
 -\beta [SI], 
 \label{eq:full_s}
 \\
 \frac{\txtd }{\txtd t} [I]
 =&
 \beta [SI] - (\gamma + \kappa)[I],
 \label{eq:full_i}
 \\
 \frac{\txtd }{\txtd t} [R]
 =&
\gamma [I] +\delta[X],
\label{eq:full_r}
 \\
 \frac{\txtd }{\txtd t} [X]
 =&
\kappa[I] - \delta[X],
\label{eq:full_x}
 \\
 \frac{\txtd }{\txtd t} [SI]
 =&
 -(\beta+\gamma+w+\kappa)[SI] + \beta[SSI] -\beta[ISI]
 \label{eq:full_si}
 \\
 \frac{\txtd }{\txtd t} [SS]
 =&
- \beta [SSI] + w\frac{[S]}{[R]+[S]} [SI]
\label{eq:full_ss}
\\
\frac{\txtd }{\txtd t}{[SR]}=&
+\gamma [SI]
+\delta [SX]
-\beta [ISR]
+ w\frac{[R]}{[R]+[S]} [SI]
% no quarantining effects
\label{eq:full_sr}
\\
\frac{\txtd }{\txtd t}{[SX]}=&
% no recovery
-\delta[SX]
-\beta [ISX]
% no rewiring effects
+ \kappa [SI]
\label{eq:full_sx}
\\
\frac{\txtd }{\txtd t}{[II]}=&
-2\gamma [II]
% no release effects
+\beta [SI] + \beta [ISI]
% no rewiring effects
-2\kappa [II]
\label{eq:full_ii}
\\
\frac{\txtd }{\txtd t}{[IR]}=&
+2\gamma [II] -\gamma [IR]
+\delta [IX]
+\beta [ISR]
% no rewiring
-\kappa [IR]
\label{eq:full_ir}
\\
\frac{\txtd }{\txtd t}{[IX]}=&
- \gamma [IX]
- \delta [IX]
+ \beta [ISX]
% no rewiring effects
+ 2\kappa [II] - \kappa [IX] 
\label{eq:full_ix}
\\
\frac{\txtd }{\txtd t}{[RR]}=&
+\gamma[IR]
+\delta[RX]
% no infection
% no rewiring
+\kappa[RX]
\label{eq:full_rr}
\\
\frac{\txtd }{\txtd t}{[RX]}=&
+\gamma[IX]
-\delta [RX] + 2\delta [XX]
% no infection
% no rewiring
+\kappa [RI]
\label{eq:full_rx}
\\
\frac{\txtd }{\txtd t}{[XX]}=&
% no recovery
-2 \delta [XX]
% no infection
% no rewiring
+ \kappa [XI]
\label{eq:full_xx}
\end{align}
We see that equations for the nodes \eqref{eq:full_s}-\eqref{eq:full_x} depend only through $[SI]$ on the equations of the links. Equation \eqref{eq:full_si} depends on itself, on $[SSI]$ and on $[ISI]$. After pair approximation there is another dependence on $[SS]$ entering, so we need \eqref{eq:full_ss}. The resulting equations are self-contained. 

If we now add further equations for the other link densities we obtain equations \eqref{eq:full_sr}-\eqref{eq:full_xx}. We also see from equations \eqref{eq:full_s}-\eqref{eq:full_xx}, that there is node and link conservation. Note that there is no dependence on the average degree. And there should not be. We could enforce it, by redefining the variables. So for instance if we have $\rho_S:= [S]/N \in [0,1]$ for the per-capita density of susceptibles in the population and $\rho_{SI}:=[SI]/N\in[0,\mu/2]$ for the average number of $[SI]$-links per node, then we could redefine $\hat{\rho_{SI}}:=[SI]/L\in[0,1]$ as an actual density (with $\hat{\rho_{SI}}=(2/\mu)\rho_{SI}$) and equation \eqref{eq:full_s} would then read $\frac{\txtd }{\txtd t} \rho_S= -\beta (\mu/2)\hat{\rho_{SI}}$. Yet, this explicit dependence on the degree is not necessary, so we have decided to employ the equations without this additional parameter.\medskip

Starting from the second-order, we can also write down the third-order equations:
\begin{align*}
\frac{\txtd }{\txtd t}{[SSI]}=&
-r[SSI]- \kappa [SSI] - w[SSI] 
+w \frac{[S]}{[S]+[R]}[SI]\frac{[SI]}{[S]} \\
&- \beta [SSI] - \beta \big[S\stackrel{I}{S}I\big] - \beta [ISSI]\\
\frac{\txtd }{\txtd t}{[ISI]}=&
-2r[ISI]- 2\kappa [ISI] - 2w[ISI] 
 - 2\beta [ISI] - \beta \big[I\stackrel{I}{S}I\big].
\end{align*}
These equations can also be written in the alternative form via pair approximation densities as follows
\begin{align*}
\frac{\txtd }{\txtd t}{\rho_{SSI}}=&
-r\rho_{SSI}- \kappa \rho_{SSI} - w\rho_{SSI} 
+w \frac{\rho_{S}}{\rho_{S}+\rho_{R}}\rho_{SI}\frac{\rho_{SI}}{\rho_{S}} - \beta \rho_{SSI} - \beta \rho_{SI}\frac{\rho_{ISI}}{\rho_{S}}- \beta \rho_{SI}\frac{\rho_{SSI}}{\rho_{S}}
\\
\frac{\txtd }{\txtd t}{\rho_{ISI}}=&
-2r\rho_{ISI}- 2\kappa \rho_{ISI} - 2w\rho_{ISI} 
 - 2\beta \rho_{ISI} - \beta \rho_{SI}\frac{\rho_{ISI}}{\rho_{S}}.
\end{align*}
Now the procedure to develop higher-order moment-closures could be pursued~\cite{KissMillerSimon} but we have not found that the resulting equations add significant practical insight in our context. Instead, we 
are going to derive to interesting variations of the equations we have analyzed.

\subsection{Adaptive SIR with infinite quarantine}

 Following~\cite{MaierBrockmann}, we consider the modification that not the links are removed but that there is a state $[X]$ which represents the quarantined state. In that state, the disease cannot be transmitted along the link. So all those links attached to that node are removed from the pool of transmittable links. Effectively this behaves like link-removal. Hence, we obtain the equations
\begin{align*}
 \frac{\txtd }{\txtd t} [S]
 &=
 -\beta [SI] - \kappa_0[S]
 \\
 \frac{\txtd }{\txtd t} [I]
 &=
 \beta [SI] - \gamma [I] - (\kappa_0+\kappa)[I]
 \\
 \frac{\txtd }{\txtd t} [R]
 &=
\gamma [I]
 \\
  \frac{\txtd }{\txtd t} [X]
 &=
\kappa_0 [S] + (\kappa_0+\kappa)[I]
 \\
 \frac{\txtd }{\txtd t} [SI]
 &=
 -(\beta+\gamma+w)[SI] + \beta[SSI] -\beta[ISI] -2\kappa_0 [SI] - \kappa [SI] 
 \\
 \frac{\txtd }{\txtd t} [SS]
 &=
- \beta [SSI] + w*\frac{[S]}{[R]+[S]} [SI] -2\kappa_0 [SS]
 \end{align*}
 This latter model has the following density representation:
  \begin{align*}
  \frac{\txtd }{\txtd t} \rho_S
  &= 
  -\beta \rho_{SI} - \kappa_0\rho_S\\
 \frac{\txtd }{\txtd t}\rho_I 
  &=
   \beta \rho_{SI} - \gamma \rho_I  - (\kappa_0+\kappa) \rho_I
   \\
     \frac{\txtd }{\txtd t} \rho_R 
  &=
    \gamma \rho_I
\\
     \frac{\txtd }{\txtd t} \rho_{SI}
&\approx
-(\beta+\gamma+w +2\kappa_0 +\kappa) \rho_{SI} +\beta \rho_{SI}\frac{ 2 \rho_{SS}-\rho_{SI}  }{\rho_S}
\\
    \frac{\txtd }{\txtd t} \rho_{SS}
&\approx
-2\beta \frac{\rho_{SI} \rho_{SS}}{\rho_S}
+w\left[\frac{\rho_S}{\rho_S+\rho_R}\right]\rho_{SI} - 2\kappa_0\rho_{SS}
 \end{align*}

\subsection{Adaptive SIR with quarantine and a return rate}

Lastly, let us consider, as before, that there is a quarantined compartment, but quarantine does not last forever and there is a return rate $\delta>0$. We still allow for both, the susceptible population as well as the infected population to be quarantined at a rate $\kappa_0$ and $\kappa_0+\kappa$ respectively. If they are healthy quarantined individuals, then they are transferred back into the susceptible compartment. If they are infected, they transition into the recovered compartment, respectively, at the aforementioned rate $\delta$. In summary, this yields the equations
\begin{align*}
 \frac{\txtd }{\txtd t} [S]
 =&
 -\beta [SI] - \kappa_0[S] +\delta[X_S]
 \\
 \frac{\txtd }{\txtd t} [I]
 =&
 \beta [SI] - \gamma [I] - (\kappa_0+\kappa)[I] 
 \\
 \frac{\txtd }{\txtd t} [R]
 =&
\gamma [I] +\delta[X_I]
 \\
 \frac{\txtd }{\txtd t} [X_S]
 =&
\kappa_0 [S] - \delta[X_S]
 \\
 \frac{\txtd }{\txtd t} [X_I]
 =&
(\kappa_0 +\kappa) [I] - \delta[X_I]
 \\
 \frac{\txtd }{\txtd t} [SI]
 =&
 -(\beta+\gamma+w)[SI] + \beta[SSI] -\beta[ISI] -2\kappa_0 [SI] - \kappa [SI] + \delta [X_{S}I] 
 \\
 \frac{\txtd }{\txtd t} [SS]
 =&
- \beta [SSI] + w*\frac{[S]}{[R]+[S]} [SI] -2\kappa_0 [SS] + \delta[X_{S}S]  \\
 \frac{\txtd }{\txtd t} [X_{S}S]
 =&
 -\delta [X_{S}S] + 2\kappa_0[SS]
 -\beta [X_{S}SI]
 \\
 \frac{\txtd }{\txtd t} [X_{S}I]
 =&
 -\delta[X_{S}I] - \gamma [X_{S}I] -(\kappa_0+\kappa)[X_{S}I] + \kappa_0 [SI]
 \end{align*}
After the usual reduction steps, we obtain a closed system of eight ODEs for the densities
 \begin{align*}
 \frac{\txtd }{\txtd t}{\rho_S}=&-\beta \rho_{SI}-\kappa_0\rho_S+ \delta \rho_{X_s}\\
\frac{\txtd }{\txtd t}{\rho_{R}}=&+
  \gamma(1-\rho_{R}-\rho_{S}-\rho_{X_s}-\rho_{X_i}) +\delta \rho_{X_i}\\
  \frac{\txtd }{\txtd t} {\rho_{X_s}}=&+
  \kappa_0\rho_S - \delta \rho_{X_s}\\
  \frac{\txtd }{\txtd t} {\rho_{X_i}}= &
  +(\kappa_0+\kappa)(1-\rho_{S}-\rho_{R}-\rho_{X_s}-\rho_{X_i}) - \delta \rho_{X_i}\\
  %&\nonumber\\
 \frac{\txtd }{\txtd t}{\rho_{SI}}=&
  -(\beta+\gamma+w+2\kappa_0+\kappa)\rho_{SI}+\beta\rho_{SI}\frac{2\rho_{SS}-\rho_{SI}}{\rho_S}+ \delta \rho_{IX_s}\\
 \frac{\txtd }{\txtd t}{\rho_{SS}}=& - 2\beta\frac{\rho_{SI}\rho_{SS}}{\rho_{S}}+w\frac{\rho_S}{\rho_S+\rho_{R}}\rho_{SI} - 2\kappa_0\rho_{SS} + \delta\rho_{SX_{s}}\\
 \frac{\txtd }{\txtd t}{\rho_{SX_{s}}}=&
  -\delta\rho_{SX_s}+2\kappa_0\rho_{SS}-\beta\rho_{SX_{s}}\frac{\rho_{SI}}{\rho_S}\\
 \frac{\txtd }{\txtd t}{\rho_{IX_{s}}}=&
  -\delta\rho_{IX_{s}}-\gamma\rho_{IX_{s}}-(\kappa_0+\kappa)\rho_{IX_{s}}+\kappa\rho_{SI}.
 \end{align*}
In future work, it could be interesting to study the differences between the slight variations of adaptive social self-distancing epidemic networks with quarantine we presented in this appendix. Yet, we conjecture that the key major effects to understand the control of the disease are already displayed by our five-dimensional main system~\eqref{eq:mfa}.

\section{Additional Comparisons between Network Simulations and the Moment Equations}
\label{sec:apB}

\begin{figure}
    \centering
    \begin{overpic}[width=0.48\linewidth]{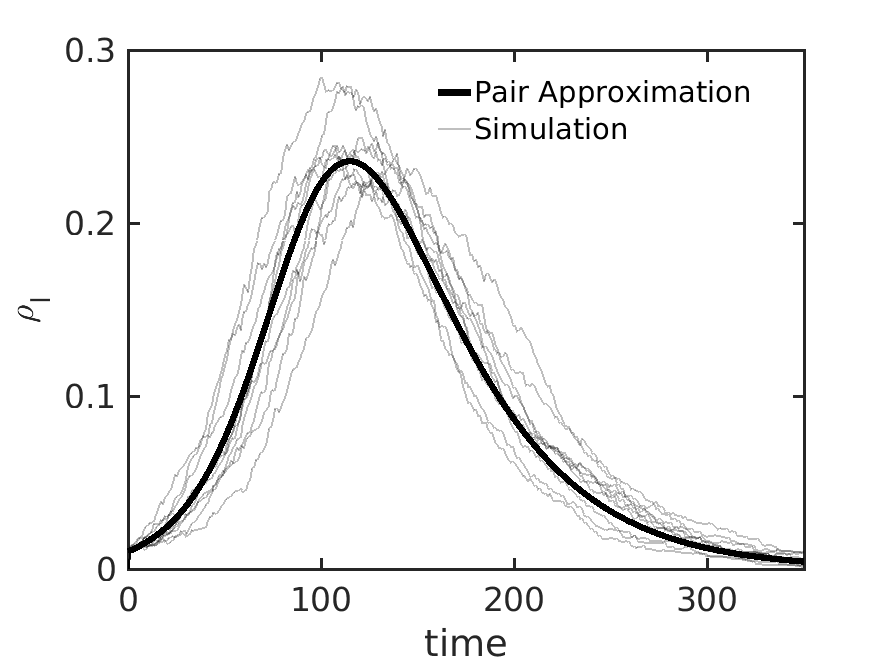}%
    \put(0,700){a)}%
    \end{overpic}
    \begin{overpic}[width=0.48\linewidth]{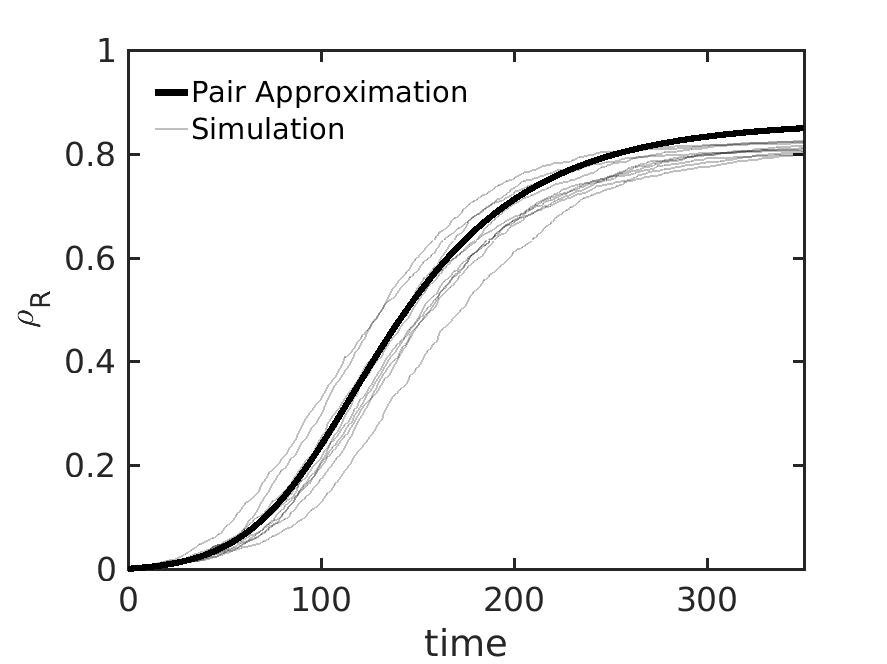}%
    \put(0,700){b)}%
    \end{overpic}
    \caption{ Sample paths for the adaptive SIRX model (thin line) and the Pair Approximation from \eqref{eq:mfa} (thick line). 
    In a) we depict the disease prevalence ($\rho_I$) and in b) we depict the cumulative size of the recovered compartment ($\rho_R$). The dynamical parameters are given by a recovery rate of $\gamma=0.025$ and an infection rate of $\beta=0.005$. The intervention parameters for the quarantine and re-wiring rates are $\kappa = 0.0025$ and $w=0$. The release rate from the quarantined compartment is $\delta = 0.001$. For the simulation we sampled from an Erd\~os-Rényi ensemble of size $N=2000$ with mean degree $\mu=15$. We initialized $1\%$ of nodes as infected $\rho_I(0) = 0.01$ and $\rho_{SI}(0) = \mu \rho_I(0)$. 
    }
    \label{fig:a1}
\end{figure}

\begin{figure}
    \centering
    \begin{overpic}[width=0.48\linewidth]{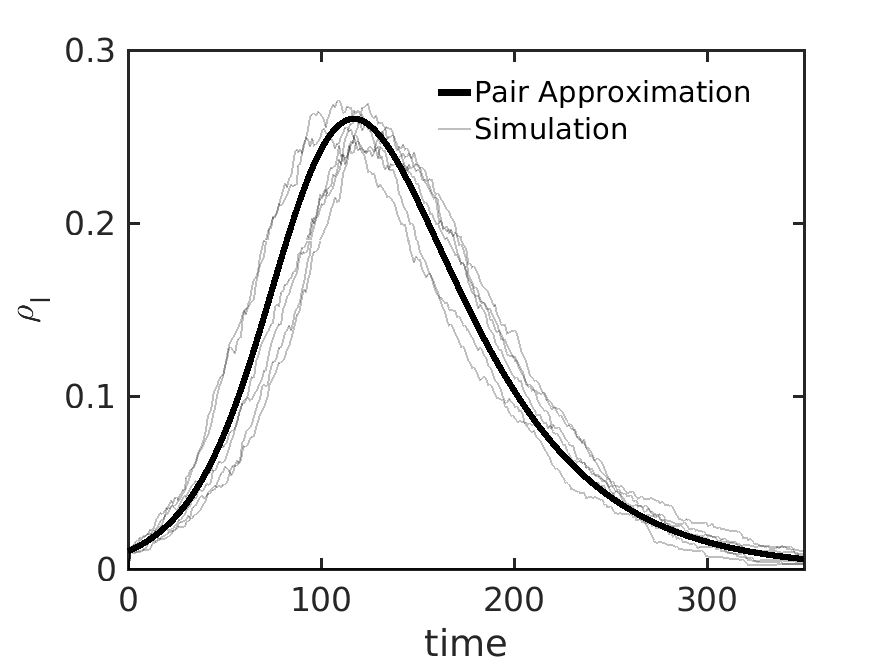}%
    \put(0,700){a)}%
    \end{overpic}
    \begin{overpic}[width=0.48\linewidth]{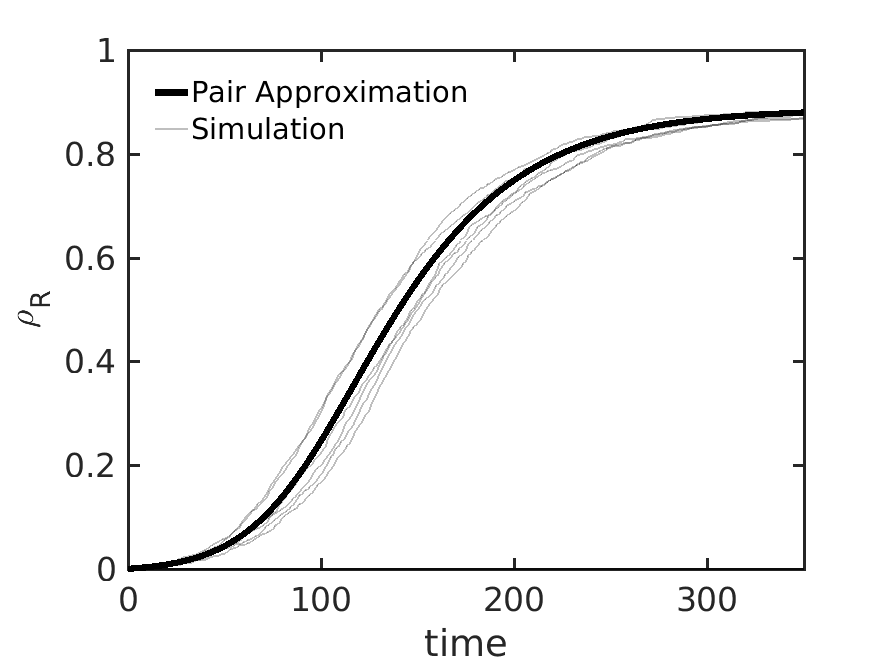}%
    \put(0,700){b)}%
    \end{overpic}
    \caption{ Sample paths for the adaptive SIRX model (thin line) and the Pair Approximation from \eqref{eq:mfa} (thick line). 
    In a) we depict the disease prevalence ($\rho_I$) and in b) we depict the cumulative size of the recovered compartment ($\rho_R$). The dynamical parameters are given by a recovery rate of $\gamma=0.025$ and an infection rate of $\beta=0.005$. The intervention parameters for the quarantine and re-wiring rates are $\kappa = 0$ and $w=0.0025$. The release rate from the quarantined compartment is $\delta = 0.001$. For the simulation we sampled from an Erd\~os-Rényi ensemble of size $N=2000$ with mean degree $\mu=15$. We initialized $1\%$ of nodes as infected $\rho_I(0) = 0.01$ and $\rho_{SI}(0) = \mu \rho_I(0)$. 
    }
    \label{fig:a2}
\end{figure}

In this appendix, we collect several additional examples of comparisons between direct network simulations and the moment-closure equations~\eqref{eq:mfa}. As outlined in the main part of the paper, we are particularly interested in the parameters $w$ controlling the re-wiring of social self-distancing and the quarantine rate $\kappa$. We have already shown a comparison in Figure~\ref{fig:1}, where $\kappa>0$ and $w>0$, which showed very good agreement between network dynamics and the moment-closure approximation. To check the robustness of the approximation, we provide here also in Figure~\ref{fig:a1} the case $w=0$ and  $\kappa>0$, while in Figure~\ref{fig:a2} we consider $\kappa=0$ and $w>0$. Even for these quite singular cases, the approximation via second-order moment-closure works extremely well.

\end{document}